\documentclass[12pt]{iopart}

\usepackage{graphicx}
\usepackage{color}

\usepackage{iopams}

\begin{document}


\title[Quantum non-equilibrium dynamics of Rydberg gases]{ Quantum non-equilibrium dynamics of Rydberg gases in the presence of dephasing noise of different strengths}

\author{Emanuele Levi, Ricardo Guti\'errez, and Igor Lesanovsky}

\address{School of Physics and Astronomy, University of Nottingham, Nottingham, NG7 2RD, United Kingdom}
\ead{ricardo.gutierrez@nottingham.ac.uk}
\vspace{10pt}
\begin{indented}
\item[]\today
\end{indented}

\begin{abstract}
In the presence of strong dephasing noise the dynamics of Rydberg gases becomes effectively classical, due to the rapid decay of quantum superpositions between atomic levels. Recently a great deal of attention has been devoted to the stochastic dynamics that emerges in that limit, revealing several interesting features, including kinetically-constrained glassy behaviour, self-similarity and aggregation effects. However, the non-equilibrium physics of these systems, in particular in the regime where coherent and dissipative processes contribute on equal footing, is yet far from being understood. To explore this we study the dynamics of a small one-dimensional Rydberg lattice gas subject to dephasing noise by numerically integrating the quantum Master equation.
 We interpolate between the coherent and the strongly dephased regime by defining a generalised concept of a blockade length.
We find indications that the main features observed in the strongly dissipative limit persist when the dissipation is not strong enough to annihilate quantum coherences at the dynamically relevant time scales. These features include the existence of a time-dependent Rydberg blockade radius, and a growth of the density of excitations which is compatible with the power-law behaviour expected in the classical limit.
\end{abstract}

%
%
%
%
%

\section{Introduction}
In the past decades experimental advances in the control of cold atomic ensembles \cite{bloch2008} have paved the way for the study of dynamical properties of quantum many-body systems \cite{greiner2002,kinoshita2006,cheneau2012}.
Among these Rydberg gases, e.g. ensembles of atoms with a valence electron excited to a state with a high principal quantum number, have been subject to intense study \cite{low2012,saffman2010}. Rydberg atoms feature strong van der Waals or dipole-dipole interactions, leading to so-called blockade effects, whereby an excited atom inhibits the laser excitation of other atoms in its vicinity \cite{jaksch2000,lukin2001}. The presence of the blockade effect has been shown to have remarkable impacts both on the statics \cite{schauss2014,schachenmayer2010,weimer2010,lesanovsky2011,levi2014,ji2011,levi2015} and the dynamics \cite{schauss2012,dudin2012,barredo2014} of Rydberg ensembles.

When studying quantum many-body systems with cold atoms the aim is typically to conduct experiments in the coherent limit, i.e. in the absence of noise. However, the experimental exploration of open quantum systems that are subject to noise has lately become very active. In Rydberg gases this dissipative regime features rich equilibrium \cite{carr2013,marcuzzi2014,malossi2014,lee2012} and out-of-equilibrium \cite{marcuzzi2015,everest2015} physics. Moreover, dissipation has been shown to be a resource for the engineering of entangled quantum states \cite{diehl2008,rao2013,lee2015}.

Recently, a number of theoretical works have focused on studying Rydberg gases in the presence of strong dephasing noise \cite{honing2013,schonleber2014,glaetzle2012,petrosyan2013,lesanovsky2013,lesanovsky2014,marcuzzi2014,gutierrez2015}. In this regime all quantum coherences are suppressed on the relevant time scale, and the time evolution of the system can be described perturbatively by means of a stochastic equation which acts only on the diagonal elements of the density matrix, i.e. classically accessible states.
This case can therefore be thought of as a classical Markovian process with configuration-dependent rates \cite{lesanovsky2013, marcuzzi2014,cai2013}.
This strongly dissipative limit displays a rich dynamics, with features such as glassiness \cite{lesanovsky2013}, aggregation effects associated with the detuning between the laser field and the atomic transition \cite{lesanovsky2014}, and scale invariance in the distribution of excitations \cite{gutierrez2015}. Some of these phenomena have also been observed experimentally, thus indicating the physical relevance of the effective dynamics \cite{valado2015,urvoy2015}.

In this paper, we investigate to what extent and in what form the results obtained in the strongly dissipative limit survive when quantum superpositions cannot be neglected on relevant dynamical time scales, and coherent and dissipative processes compete. For this purpose, we integrate numerically the quantum Master equation of the system for different dissipation strengths. For computational reasons, we consider a chain of $N=8$ atoms. This is far from the large number of particles which can be tackled in the classical limit but nevertheless such system sizes are of relevance for current experimental setups that explore Rydberg ensembles in microtrap arrays \cite{labuhn2015}.

In order to be able to systematically explore the transition from the classical to the quantum limit we start by introducing a generalised blockade radius. This allows us to ensure that the only relevant length scale in the non-equilibrium dynamics is kept constant as the dissipation strength is varied. We then explore the growth of the density of Rydberg excitations and the dynamics of Rydberg-blockade effects as seen in the behaviour of two-point correlations. As a main result we find indications that there is a regime in which the system displays a power-law growth of the density of excitations (as predicted in \cite{lesanovsky2013} for the classical limit) even if the dissipation rate is only moderately high, or even weak compared to the driving.

We would like to remark that the interplay between classical and quantum dynamics in Rydberg ensembles was also explored in a recent paper by Mattioli {\it et al.} \cite{mattioli2015} in a slightly different context. Here a system of three-level atoms that include a short-lived excited state with a lower energy than the Rydberg state was investigated. Such system allows for the non-equilibrium dynamics to be tuned from the classical to the quantum regime by varying the strength of the laser driving to the intermediate level. While both papers have some conceptual similarities, our perspective is somewhat different, as we aim to explore the influence of increasing quantum effects starting from a classical setting whose dynamics has been thoroughly investigated in the past.

The paper is organised as follows: in Section 2 we provide details on the numerical methods and discuss the definition of the blockade radius as well as its dependence on the system parameters. In Section 3 we explore the dynamics of the excitation density and the density-density correlation function for different dissipation strengths. Section 4 contains a summary and discussion of the results as well as an outlook.

\section{Model, simulations, and generalised blockade radius}

\subsection{The dissipative Rydberg gas}
We consider an open one dimensional lattice gas of $N=8$ atoms subject to a laser field that resonantly couples the ground state $|g\rangle$ to a Rydberg state $|e\rangle$ with Rabi frequency $\Omega$. We consider $\hbar=1$ and express all the physical parameters in units of a reference frequency $\Omega_0$. Equivalently time is given in units of $1/\Omega_0$. The interactions between atoms in the Rydberg state are assumed to be of van der Waals type and the interaction energy between these atoms at positions ${\bf r}_k$ and ${\bf r}_m$ is $V_{km} = C_6/|{\bf r}_k - {\bf r}_m|^6$. In the rotating wave approximation the Hamiltonian of the system is then given by
\begin{equation}
H = \Omega \sum_k \sigma_x^{(k)} + \sum_{k < m} V_{km}\, n^{(k)} n^{(m)}.
\label{h}
\end{equation}
Here $\sigma_{x}^{(k)}=\left|e\right>_k\!\left<g\right| + \left|g\right>_k\!\left<e\right|$, and $n^{(k)}=\left|e\right>_k\!\left<e\right|$.
We introduce dephasing noise by considering our system as weakly coupled to a memoryless bath, such that the dynamics can be described by the Lindblad Master equation \cite{lindblad,breuer2002}
\begin{equation}
\partial_t \rho = -i[H,\rho] + \gamma \sum_{k=1}^N\left(n^{(k)} \rho\, n^{(k)} - \frac{1}{2} \left\{n^{(k)},\rho\right\}\right).
\label{l0}
\end{equation}
The form of the Master equation (\ref{l0}) has been derived microscopically for cold atoms held in the lowest band of an optical lattice (see e.g. \cite{pichler2010}).   
Note, that we do not consider here radiative decay of atomic excitations which inevitably occurs in Rydberg systems. We perform this simplification in order not to further obscure the dynamical features we are interested in exploring. They are already challenging to extract due to the small size of the system under consideration.

In order to solve Eq. (\ref{l0}) one needs to integrate a system of $2^{N-1}(2^N+1)$ complex ordinary differential equations, which is a challenging numerical task even for relatively small systems.
In this paper we use both a numerical integration, and a quantum jumps Monte Carlo (QJMC) unraveling of Eq. (\ref{l0}) to explore the dynamics \cite{molmer1993,plenio1998}. 

In the regime of strong dissipation $\gamma \gg \Omega$ the description of the system simplifies and the dynamics of the Rydberg excitations is governed by a classical stochastic equation. This can be formulated perturbatively at second order in $\Omega$ as
\begin{equation}
\partial_t \mu = \frac{4\Omega^2}{\gamma} \sum_k \Gamma_k \left[\sigma_{x}^{(k)} \mu \sigma_{x}^{(k)} -  \mu \right],
\label{effective}
\end{equation}
where  $\mu = \textrm{diag}(\rho)$ \cite{lesanovsky2013,marcuzzi2014} is diagonal in the Fock basis.  A change in the state of site $k$, whether an excitation $|g\rangle \to |e\rangle$ or a de-excitation $|e\rangle \to |g\rangle$, occurs with an (operator-valued) rate $\Gamma_k$, such that
\begin{equation}
\Gamma_k^{-1} = 1 + \left[ R^6 \sum_m  \frac{n^{(m)}}{|{\bf r}_k - {\bf r}_m|^6}\right]^2.
\label{rates}
\end{equation}
Due to the structure of this rate, state changes of atoms proceed much more slowly in the vicinity of excited atoms than far away from them, which results in a strongly correlated dynamics. The relevant distance at which the crossover between slow and fast dynamics is taking place is given by $R \equiv \left(2\, C_6/\gamma \right)^{1/6}$ which we refer to as the blockade radius for the effective strongly dissipative dynamics.

The emergence of classical dynamics in the strongly dissipative case can be better understood with an illustrative example based on QJMC.
\begin{figure}[t]
\begin{center}
\hspace{2cm}
\includegraphics[scale=0.190]{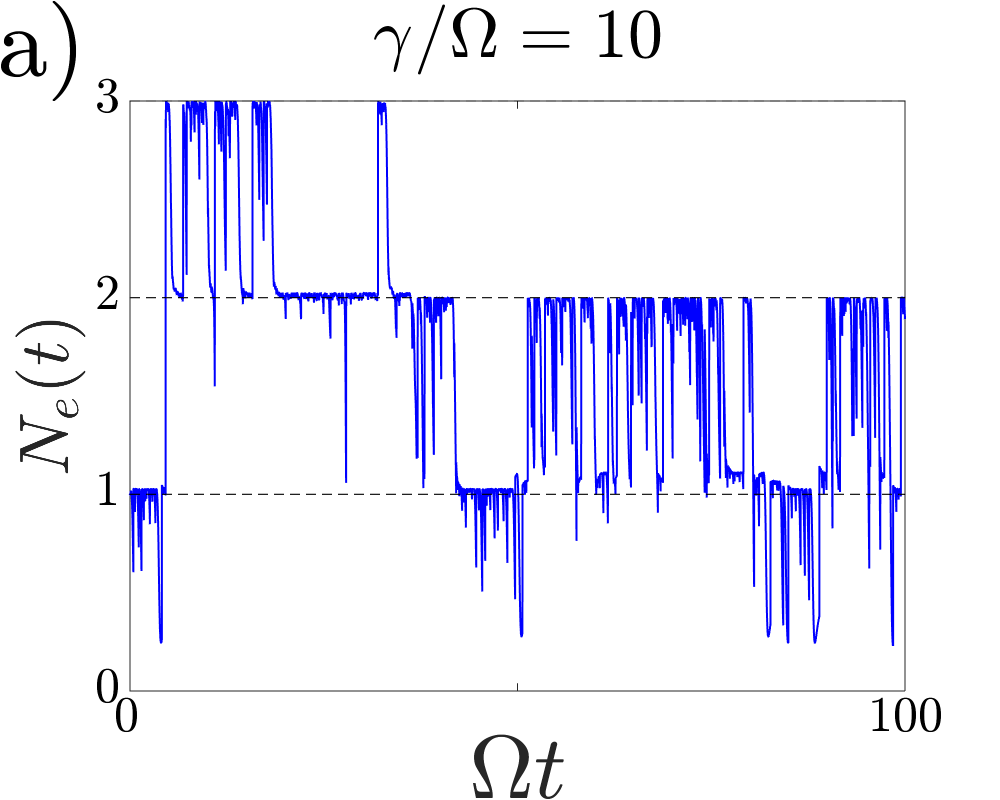}
\includegraphics[scale=0.190]{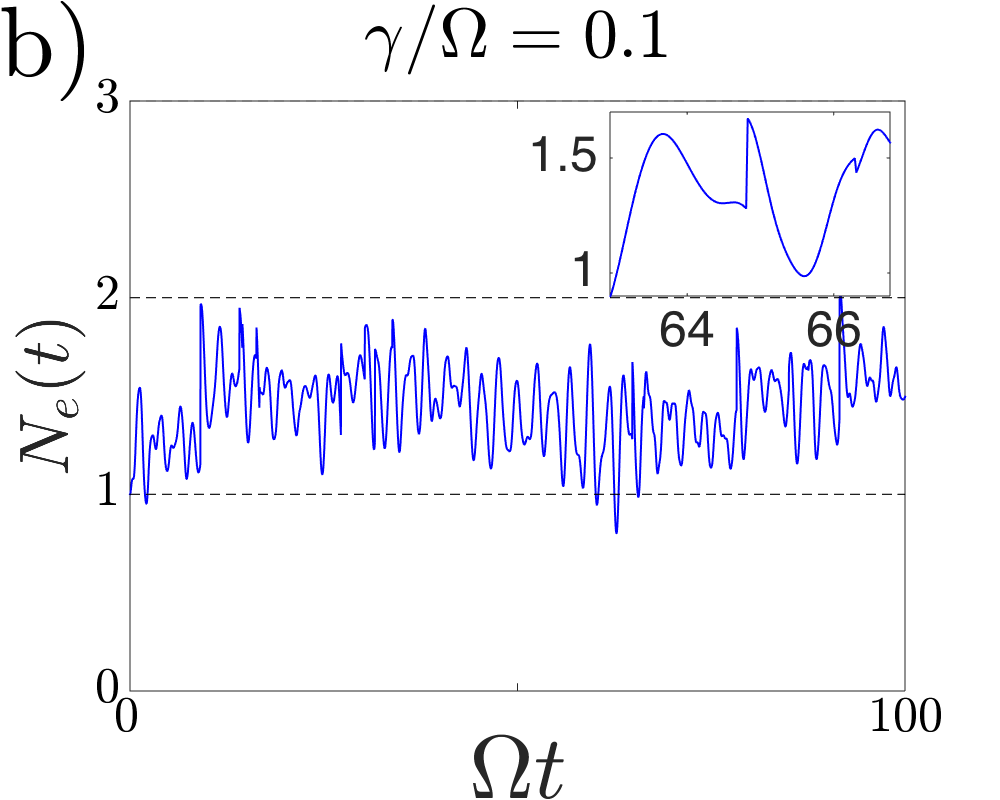}
\end{center}
\caption{ {\sf \bf Expectation value of the total number operator as a function of time for $\gamma/\Omega = 10$ and $\gamma/\Omega = 0.1$.} (a) Single QJMC trajectory for $8$ atoms starting from the state $|ggggeggg\rangle$ for $\Omega = 0.01$ and $\gamma = 0.1$ ($\gamma/\Omega = 10$). (b) Analogous curve for $\Omega = 0.020869$ and $\gamma = 0.0020869$ ($\gamma/\Omega = 0.1$). The inset shows a portion of the trajectory where it becomes clear that quantum jumps occasionally disrupt the smooth evolution of the system.}
\label{fig1}
\end{figure}
In Figure \ref{fig1} we show the expectation value of the total number operator $N_e(t) =  \sum_{i=1}^N \langle \psi(t)|n_i|\psi(t)\rangle$ corresponding to one trajectory in both the strongly dissipative (a) and the weakly dissipative (b) limit. In the presence of strong dissipation the system is subject to frequent jumps, which can be regarded as quantum measurements of the observable $n^{(k)}$ that project the wave function onto the subspace where an excitation is present on site $k$. This allows the state to be described for the vast majority of times by a statistical mixture of Fock states, that form the diagonal of the density matrix and display an integer number of excitations. In the presence of weak dissipation, although jumps still occur, the large time between jumps allow the system to create quantum superpositions which cannot be accounted for by considering only the diagonal elements of the density matrix.

\subsection{Blockade radius in the crossover region between quantum and classical regime}\label{sec:blockade_radius}

Our aim is to have a closer look at the dynamics away from the classical limit. In the extreme case when the system evolves fully coherently its dynamics is subjected to the well-know dipole blockade. Here interaction induced energy shifts inhibit the coherent excitation of Rydberg states in the vicinity of excited atoms \cite{jaksch2000,lukin2001}. In this context, the blockade radius is conventionally defined to be the distance at which the interaction strength is equal to the Rabi frequency, $r_\mathrm{b} = \left(C_6/\Omega\right)^{1/6}$. In the strongly dissipative limit, on the other hand, the relevant blockade radius is given by $R = \left(2\, C_6/\gamma\right)^{1/6}$ which depends solely on the dissipation rate \cite{lesanovsky2013}. In order to systematically explore the crossover from the strongly dissipative regime to the fully coherent limit we need to study the dynamics for different ratios $\gamma/\Omega$. We thus require a {\it generalised blockade radius} $r_\mathrm{G}$ that interpolates between the two limiting cases:  $r_\mathrm{G} \sim r_\mathrm{b}$ for $\gamma/\Omega =0$, and $r_\mathrm{G} \sim R$ as $\gamma/\Omega \to \infty$.

\begin{figure}[t]
\begin{center}
\hspace{2cm}
\includegraphics[scale=0.250]{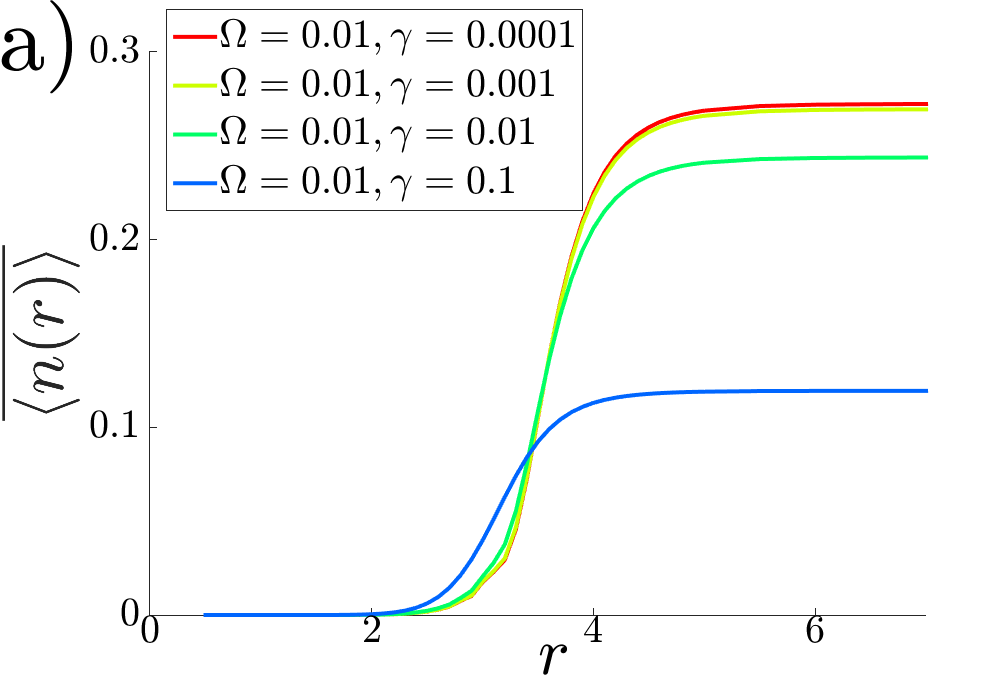}
\includegraphics[scale=0.250]{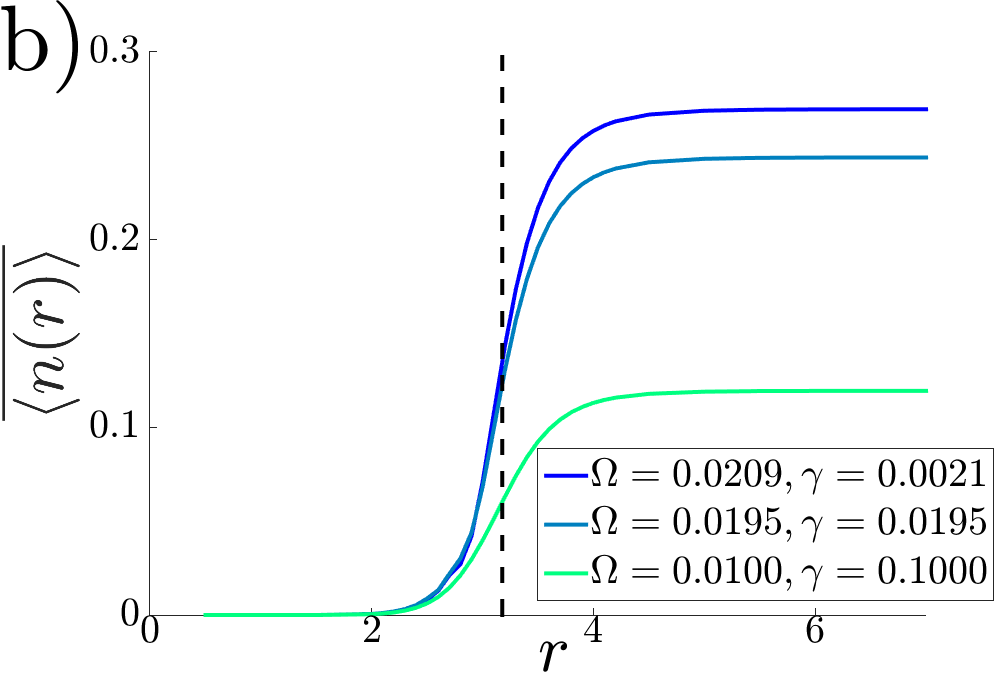}\\
\vspace{0.8cm}
\hspace{2cm}\includegraphics[scale=0.280]{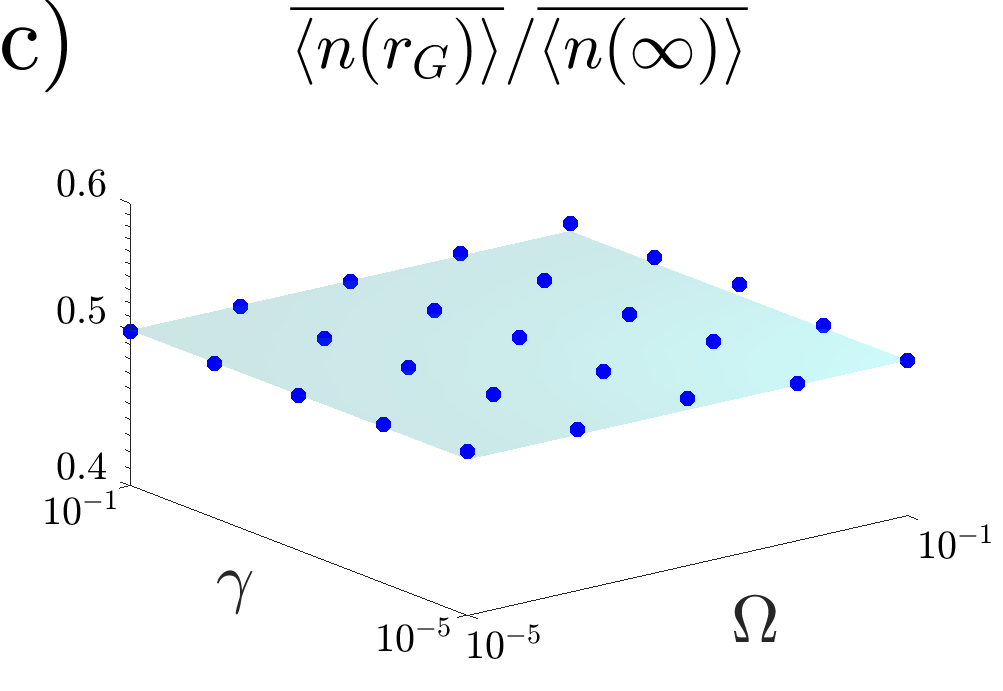}
\end{center}
\caption{ {\sf \bf Blockade radius in the crossover region between fully coherent and classical regime.} (a) Growth of the average of the excitation number operator of an atom $\overline{\langle n(r) \rangle}$ as a function of the distance $r$ from another, permanently excited atom, for $\Omega = 0.01$ and $\gamma = 0.0001, 0.001, 0.01$ and $0.1$.  (b) Average excitation number $\overline{\langle n(r) \rangle}$ for $\gamma/\Omega = 10, 1$ and $0.1$ with the same $r_\mathrm{G}$ that results from considering the case $\gamma = 0.1$ and $\Omega = 0.01$ using the interpolation formula in Eq. (\ref{interp}). The resulting parameter values are indicated in the legend, and the position of $r_\mathrm{G}$ is shown by a dashed vertical line. (c)  $\overline{\langle n(r_\mathrm{G}) \rangle}/\overline{\langle n(\infty) \rangle}$ with $r_\mathrm{G}$ values based on Eq. (\ref{interp}) for many $(\gamma,\Omega)$ pairs.}
\label{fig2}
\end{figure}

To find a suitable interpolation we study the dynamics of two interacting atoms in the presence of dissipation, which contains the essential physics of the problem.
One atom (assumed to be at the origin of the coordinate system) is permanently excited, while the other one is located at a distance $r$. Its state $\rho_1$ evolves in time according to the Master equation (we make the parametric dependence on $r$ explicit)
\begin{equation}
\partial_t \rho_1(t,r) = - i [H_1(r), \rho_1(t,r)] + \gamma \left(n \rho_1(t,r)\, n - \frac{1}{2} \left\{n,\rho_1(t,r)\right\}\right),
\label{qme2b}
\end{equation}
with the Hamiltonian
\begin{equation}
H_1(r) = \Omega\, \sigma_x + C_6 \frac{n}{r^6}.
\label{h2b}
\end{equation}
We proceed by calculating the expectation value of the Rydberg number operator $\langle n(t,r) \rangle = \textrm{Tr}[n\,\rho_1(t,r)]$ starting from the initial state $\rho_1(0,r) = |g\rangle \langle g|$.  The solution is integrated over a Rabi cycle by which we obtain the average $\overline{\langle n(r) \rangle} = \Omega \int_0^{1/\Omega} \langle n(t,r) \rangle\, dt$. For any given values of $\gamma$ and $\Omega$, $\overline{\langle n(r) \rangle}$ is seen to approach zero for sufficiently small $r$, grow suddenly at some point as $r$ is increased, and saturate for large $r$. As an illustration we show $\overline{\langle n(r) \rangle}$ for $\Omega = 0.01$ and different values of $\gamma$ in Fig. \ref{fig2} (a), where $C_6 = 80$. For sufficiently small $\gamma$ we see that the form of  $\overline{\langle n(r) \rangle}$  hardly changes, as the blockade effect is dominated by $\Omega$, and the blockade radius is given by $r_\mathrm{b}$.

We now define the generalised blockade radius $r_\mathrm{G}$ as the distance $r$ at which the growth of $\overline{\langle n(r) \rangle}$ has reached half of the saturation value, $\overline{\langle n(r_\mathrm{G}) \rangle}/\overline{\langle n(\infty) \rangle} = 1/2$. We were unable to derive a useful analytical expression for $r_\mathrm{G}$ directly from the equations of motion (\ref{qme2b}), but found that the formula
\begin{equation}
r_\mathrm{G} = \left(\frac{2\, C_6}{[a\, \Omega^p + \gamma^p]^{1/p}}\right)^{1/6}
\label{interp}
\end{equation}
with $a = 11.1$ and $p = 6/5$ provides an accurate interpolation between numerically calculated values for $r_\mathrm{G}$.

We can now use Eq. (\ref{interp}) to obtain pairs ($\gamma$, $\Omega$) which keep the blockade radius $r_\mathrm{G}$ constant but allow us to systematically move away from the classical limit. For example, if we take as a reference the blockade radius obtained from $\gamma = 0.1$ and $\Omega = 0.01$ ($\gamma/\Omega = 10$), and extract the parameter values that give the same $r_\mathrm{G}$ for $\gamma/\Omega = 1$ and $0.1$, we obtain the curves shown in Fig. \ref{fig2} (b). Here the position of $r_\mathrm{G}$ is shown by a dashed vertical line. To validate this approach more generally, we calculate $\overline{\langle n(r_\mathrm{G}) \rangle}/\overline{\langle n(\infty) \rangle}$, where $r_\mathrm{G}$ changes from one case to another according to the interpolation formula in Eq. (\ref{interp}), across four orders of magnitude in both $\gamma$ and $\Omega$. This calculation encompasses highly dissipative regimes, such as that given by  $\gamma/\Omega = 10^4$, and very coherent cases, such as $\gamma/\Omega = 10^{-4}$. The results are shown in Fig. \ref{fig2} (c). The deviation from $\overline{\langle n(r_\mathrm{G}) \rangle}/\overline{\langle n(\infty) \rangle} = 0.5$  never exceeds $1.6\%$ and is generally substantially smaller. This shows that the interpolation given by the phenomenological expression in Eq. (\ref{interp}) indeed captures the blockade radius (at least in a system of two atoms) for varying degrees of dissipation.

\section{Results}
\begin{figure}[t]
\begin{center}
\hspace{1.2cm}
\includegraphics[scale=0.25]{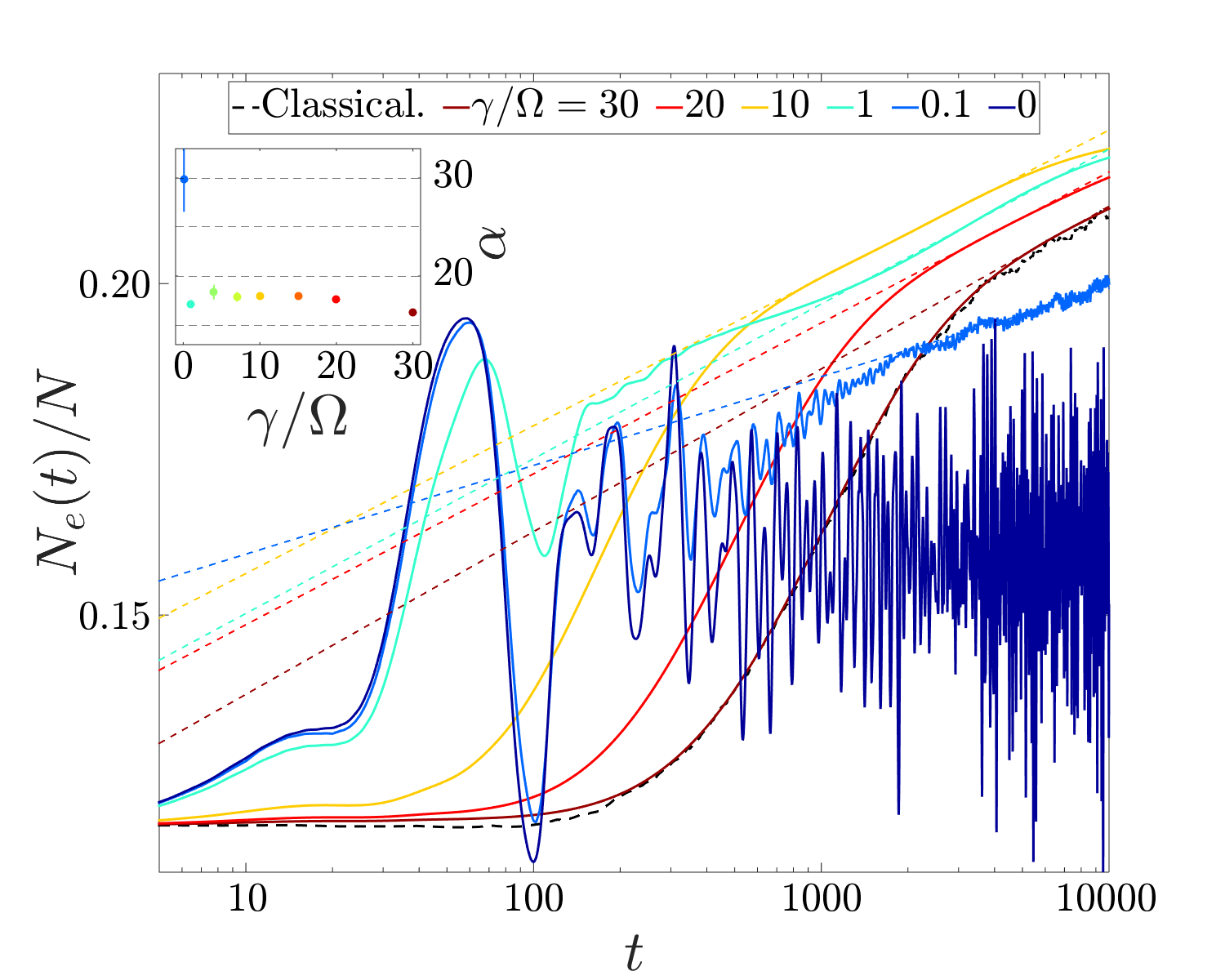}
\end{center}
\caption{ {\sf \bf Time evolution of the density of excitations.} In the main panel the dynamics of the density of excitations $N_e(t)/N$ is plotted as a function of time for various $\gamma/\Omega$ ratios. The curve $\gamma/\Omega=0.1$ is obtained with a QJMC simulation and it is averaged over 1000 trajectories. The curves $\gamma/\Omega=0,1,10,20,30$ are obtained from the integration of the quantum Master equation. Dashed lines with the same colour coding show power-law fits to these curves, assuming $N_e(t)\sim t^{1/\alpha}$, for comparatively long times, where such an algebraic growth is observed. In the case $\gamma/\Omega=30$ we draw for comparison the results obtained with a numerical integration of Eq. (\ref{effective}) by means of a kinetic Monte Carlo approach averaged over $10000$ classical trajectories. The inset shows the dependence of $\alpha$ on the ratio $\gamma/\Omega$ as extracted from the power-law fits.}
\label{fig3}
\end{figure}

In order to shed light on the interplay between the quantum driving and the dissipation we study numerically the dissipative dynamics of an open chain of $N=8$ atoms initialised in the pure state $\left| \psi_0\right\rangle=\left|ggggeggg\right\rangle$, i.e. with the $5$th atom being excited. We consider a van der Waals coefficient of $C_6 = 160$. We keep the generalised blockade radius Eq. (\ref{interp}) constant at $r_\mathrm{G}=3.566$ (which stems from choosing $\gamma = 0.1$ and $\Omega = 0.01$ for the case $\gamma/\Omega = 10$), and we consider different ratios of $\gamma/\Omega$, as shown in Fig. \ref{fig3}. In this Figure we report the quantum average of the density of excitations operator $N_e(t)=\mathrm{Tr}\left[\left(\sum_{k=1}^N n_k\right)\rho(t)\right]$ as a function of time for each case considered.
In the coherent case ($\gamma=0$) the density of excitations oscillates strongly for all times which is a consequence of the coherent nature of the evolution and the small system size. Larger systems would display relaxation towards a stationary state value \cite{weimer2008,lesanovsky2010,ates2012,cohen2016}. In the dissipative cases, the density of excitations grows towards a stationary value corresponding to the stationary density matrix and excitation number of $\rho_\mathrm{st} \equiv 2^{-N} \otimes_k \mathbb{I}_k$ and $N_e^\mathrm{st}=N/2$, respectively, regardless of the system size. Before reaching the stationary state and for very strong dissipation, i.e. in the classical limit, the system is known to display an algebraic regime where the number of excitations grows as $N_e(t)\sim t^{1/\alpha}$ \cite{lesanovsky2013,marcuzzi2014, gutierrez2015}. This power law behaviour originates from the way the system is approaching its stationary state. The blockade excitations tend to be created as far away from each other as possible which leads to a hierarchical and in fact self-similar relaxation dynamics. As a result the system``looks the same" at different times, with only the length scale changing continuously \cite{gutierrez2015}. The exponent $\alpha$ which characterises the density growth depends on the dimensions of space and the power of the interaction. In one dimension and for van der Waals interactions one expects $\alpha\approx 13$.

For our system this classical limit is approximately reached at $\gamma/\Omega=30$ as the comparison between the brown and the black dashed line in Fig. \ref{fig3} suggests. Note, however, that the regime of power law growth becomes more pronounced the larger $R$ \cite{lesanovsky2013}. The small systems size studied here restrict us to a relatively small value of $R$ and hence only the onset of the power law regime is visible. The observed exponent is approximately $16$ and thus reasonably close to the expected value given the inevitable finite size effects and the small region of interpolation. More interesting is the suggestion that the power law growth of the excitation density persists outside the classical limit, including the case with $\gamma=\Omega$ and even $\gamma\ll\Omega$, as can be seen in Fig. \ref{fig3}. The corresponding inverse exponent $\alpha$ appears to approach its classical value $\alpha\approx 13$ as the ratio $\gamma/\Omega$ is increased, as shown in the inset. This suggests that the hierarchical relaxation which, in the classical case \cite{lesanovsky2013,gutierrez2015}, was identified to underly the power law behaviour persists in the quantum regime.
We would like to stress that the values of $\alpha$ displayed in the caption are obtained with a fit over less than two decades, which is most certainly not sufficient to obtain accurate estimates. 
Better estimates could be achieved by considerably increasing the system size, and the time range considered. 
Both these cases are unfeasible with current numerical methods, but they should be accessible in current experiments. 
In the Appendix we include a study of the excitation dynamics for smaller systems which display features that are qualitatively similar, indicating that the case of $N=8$ atoms is not special in any way. However, we would like to emphasise that a detailed and unambiguous study of the algebraic growth inevitably requires the consideration of larger systems which are currently out of reach.
\begin{figure}[t]
\begin{center}
\hspace{1.4cm}
\includegraphics[scale=0.25]{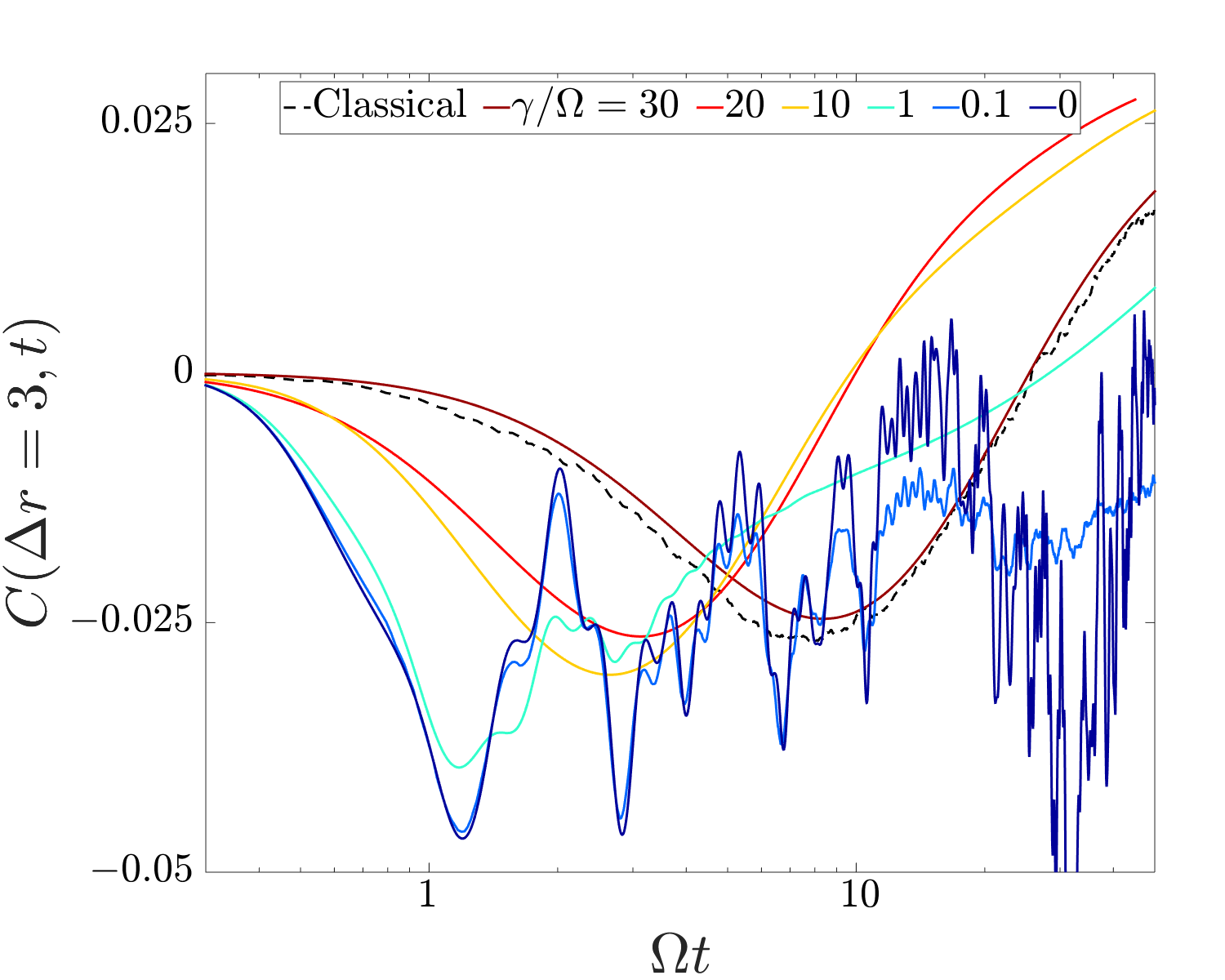}
\end{center}
\caption{ {\sf \bf Rydberg blockade dynamics as reflected in the two-point correlation function.}
Dynamics of the two-point correlation $C(\Delta r = 3,t)=\frac{1}{2} \sum_{\Delta r = \pm 3}[ \langle n_5(t) n_{5+\Delta r}(t)\rangle - \langle n_5(t)\rangle \langle n_{5+\Delta r}(t)\rangle]$ (see text for the explanation) as a function of time for various ratios of $\gamma/\Omega$. The details regarding the numerical simulations are as provided in the caption of Fig. \ref{fig3}.}
\label{fig4}
\end{figure}

Next, we study the dynamics of the blockade effect. To this end, we compute the correlation function between the atom at position $k=5$ that is initially excited and the two neighbours at distance $\Delta r = 3$: $C(\Delta r = 3,t) = \frac{1}{2} \sum_{\Delta r = \pm 3}[ \langle n_5(t) n_{5+\Delta r}(t)\rangle - \langle n_5(t)\rangle \langle n_{5+\Delta r}(t)\rangle]$. Notice that with the used initial state the atoms at distance $\Delta r = 3$ are expected to be initially blockaded, since $r_\mathrm{G} > 3$. The $C(\Delta r = 3,t)$ correlation function is shown in Fig. \ref{fig4} for the same parameter choices that were used in Fig. \ref{fig3}. To improve the representation of the results, we rescale the time by the typical time of a Rabi oscillation $1/\Omega$. We see that in the fully coherent case the correlation function oscillates strongly and excitations are anti-correlated for almost all times. This is expected since the blockade is enforced due to energetic reasons that prevent the simultaneous excitation of nearby atoms.

In the dissipative case, however, we observe that after an initial period of anti-correlation $C(\Delta r = 3,t)$ drifts towards positive values. The reason is that the dissipative system allows transitions which are ``energetically forbidden" in the closed system. In fact this is necessary to ultimately reach the uncorrelated stationary state in which excited atoms can appear at any distance. As a consequence the blockade is dynamical in the sense that the effective blockade radius is shrinking with time \cite{gutierrez2015}. Note that, as in Fig \ref{fig3}, for sufficiently large ratio $\gamma/\Omega$ the results are reasonably well reproduced by classical Monte Carlo simulations (see the case $\gamma/\Omega = 30$).

\begin{figure}[t]
\begin{center}
\hspace{1.8cm}
\includegraphics[scale=0.28]{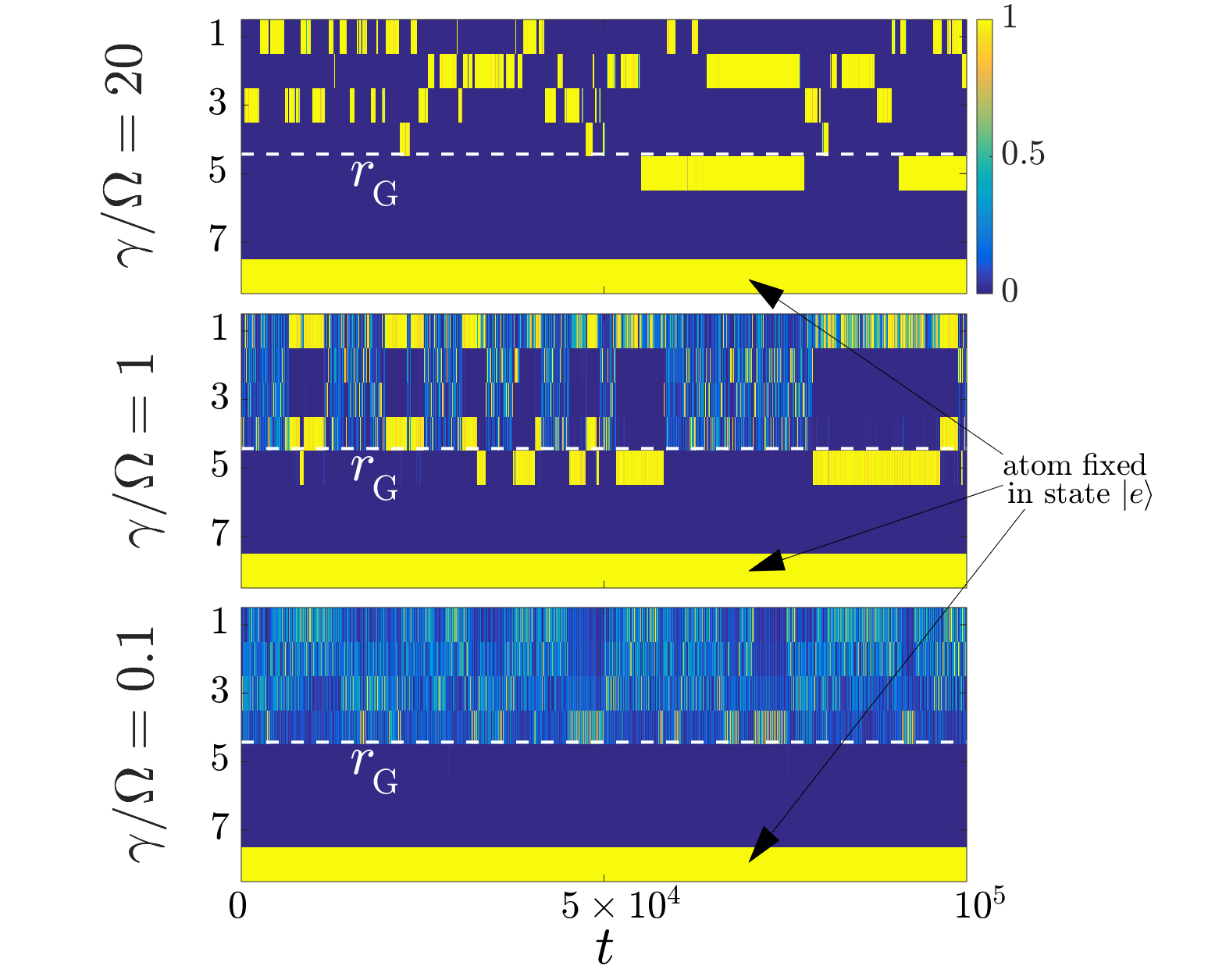}
\end{center}
\caption{ {\sf \bf Individual QJMC trajectories for different dissipation strengths.}
Individual QJMC trajectories, showing $\langle n_k\rangle$ as a function of time for various ratios of $\gamma/\Omega$. The dashed white line indicates the position of the blockade radius $r_\mathrm{G}$ with respect to atom $8$ whose state is fixed to $|e\rangle$.}
\label{fig5}
\end{figure}

To further elucidate the underlying excitation dynamics we inspect individual QJMC trajectories in a slightly modified setting. To this end we fix the atom at position $8$ in the excited state. This simplifies the visualisation of competing classical and quantum effects and sheds light on the blockade mechanism. Three QJMC trajectories, for $\gamma/\Omega = 20, 1$ and $0.1$, are shown in Fig. \ref{fig5}. We highlight with a dashed white line the position of the generalised blockade radius $r_\mathrm{G}$ with respect to the fixed excitation at site $8$.

We see that, in agreement with the illustration provided in Fig. \ref{fig1}, larger dissipation leads to a more classical dynamics, while for weak dissipation the presence of quantum superpositions becomes more evidently manifest. In all cases the first excitations emerge beyond the blockade radius $r_\mathrm{G}$, regardless of the strength of the dissipations. At later times, however, individual excitations can penetrate into the blockaded domain which is ``energetically forbidden" and thus can only be accessed due to the dissipative dynamics and not by the quantum evolution. Note, that also in the case of $\gamma/\Omega=0.1$ (bottom panel) such events can take place. However, they are rarer and on average take longer to occur.

\section{Summary and Discussion}
The aim of this paper was to investigate the interplay between quantum driving and dissipation in a Rydberg lattice gas. In particular, we were interested in understanding the dynamics of this system using the strongly dissipative limit as starting point. Here the evolution of the system is governed by effectively classical rate equations which predict interesting features such as an algebraic growth of the excitation density (with exponent $1/\alpha$) and a dynamically shrinking blockade radius. Our results suggest that these features persist until deep into the quantum regime.
This can be qualitatively understood in the following way. Even deep in the quantum regime (i.e. $\Omega\gg \gamma$) the stationary state is the completely mixed state. 
This means that there is a timescale beyond which the density of excitations $\nu$ is sufficiently large so that the interaction of an atom with the ``background'' will shift its coherent excitation out of resonance. 
Therefore the approach to the steady state must eventually proceed through incoherent jumps in order for the system to reach its dense stationary state.
This may provide an explanation on why the classical power-law growth can also be observed in the quantum case.

To illustrate this more formally we focus on the Hamiltonian associated to a given atom
\begin{equation}
 \tilde{H}^{(k)}=\Omega \sigma_x^{(k)}+2C_6 \nu^6 n^{(k)}.
\end{equation}
Here we assume a homogeneous background of excitations with average density $\nu=\sum_j n^{(j)}/N$, and we truncate the interactions to the neighbouring sites.
The quantum state evolves as
\begin{equation}
\partial_t \tilde{\rho}^{(k)}(t) = - i [\tilde{H}^{(k)}, \tilde{\rho}^{(k)}] + \gamma \left(n^{(k)} \tilde{\rho}^{(k)}\, n^{(k)} - \frac{1}{2} \left\{n^{(k)},\tilde{\rho}^{(k)}\right\}\right).
\label{qme2btilde}
\end{equation}
Denoting the expectation value of an operator $\tilde{O}^{(k)}=\Tr\left[\tilde{\rho}^{(k)}(t)O^{(k)}\right]$ the equations of motion for $\sigma_{x,y}^{(k)},n^{(k)}$ are
\begin{eqnarray}
  \dot{\tilde{n}}^{(k)}&=\Omega \tilde{\sigma}_x^{(k)},\nonumber\\
  \dot{\tilde{\sigma}}_x^{(k)}&=-\frac{\gamma}{2}\tilde{\sigma}_x^{(k)}-2C_6\nu^6\tilde{\sigma}_y^{(k)},\nonumber\\
  \dot{\tilde{\sigma}}_y^{(k)}&=2\Omega\left(1-2\tilde{n}^{(k)}\right)-\frac{\gamma}{2}\tilde{\sigma}_y^{(k)}+2C_6\nu^6\tilde{\sigma}_x^{(k)}.
\label{eq:eqmotion}
\end{eqnarray}
For sufficiently high densities $\nu$, the interaction is so strong that the coherent excitation is off-resonant and thus we can set $\dot{\tilde{\sigma}}_{x,y}^{(k)}\simeq0$. This means that the creation of coherences is strongly suppressed. 
In this regime we can solve Eq.~(\ref{eq:eqmotion}) for $\tilde{n}^{(k)}$, and we can extract the evolution of the average density of excitations by summing over all sites
\begin{equation}
 \dot{\nu}=\frac{4\Omega^2}{\gamma}\frac{2}{1+4\left(\frac{2C_6}{\gamma}\nu^6\right)^2}\left(\frac{1}{2}-\nu\right).
 \label{eq:final}
\end{equation}
This equation is similar in form to the rate equation derived in \cite{lesanovsky2013}. 
For strong interactions and intermediate times ($\nu \ll 1/2$), we can solve Eq.~(\ref{eq:final}) obtaining $\nu\sim t^{1/13}$.
This suggests the emergence of a regime of algebraic density growth as is observed in our simulations. Notice that the value of the exponent $\alpha=13$ is independent of the value of $\gamma/\Omega$, suggesting that the dependence shown in the inset of Fig.~\ref{fig3} might indeed stem from finite size effects.

Clearly, the current study suffers from the fact that it is purely numerical and that it focusses on small system sizes. Establishing a conclusive characterisation of the dynamical features and in particular their dependence on competing classical noise (dissipation) and quantum fluctuations (coherent driving) is therefore not possible. Nevertheless, we believe that the present study is sufficiently interesting to motivate further investigations. The proposed generalised blockade radius, which proved successful in fitting our two--atoms results, could find an application in such future studies. It would be interesting to see whether this empirical result can be put on more solid theoretical grounds. Experiments conducted in the classical limit have established the validity of the rate equation description \cite{urvoy2015,valado2015}. Further experiments could probe the quantum regime in order to study the algebraic growth of excitations and the evolution of the correlation function. From the theoretical side one possibility to make progress is to resort to a field theoretical treatment as e.g. conducted in Refs. \cite{maghrebi2015,marcuzzi2016} in order to shed light on the interplay between coherent excitations and noise.

\section*{Acknowledgements}
The research leading to these results has received funding from the European Research Council under the European Union's Seventh Framework Programme (FP/2007-2013) and ERC Grant Agreement No. 335266 (ESCQUMA), the EU-FET grant HAIRS, 612862 and from the University of Nottingham. Further funding was received through the H2020-FETPROACT-2014 Grant No. 640378 (RYSQ). We also acknowledge financial support from EPSRC Grant No. EP/J009776/1. Our work has benefited from the computational resources and assistance provided by the University of Nottingham High Performance Computing service.

\appendix
\section{}
\begin{figure}[h]
\begin{center}
\hspace{1.8cm}
\includegraphics[scale=0.28]{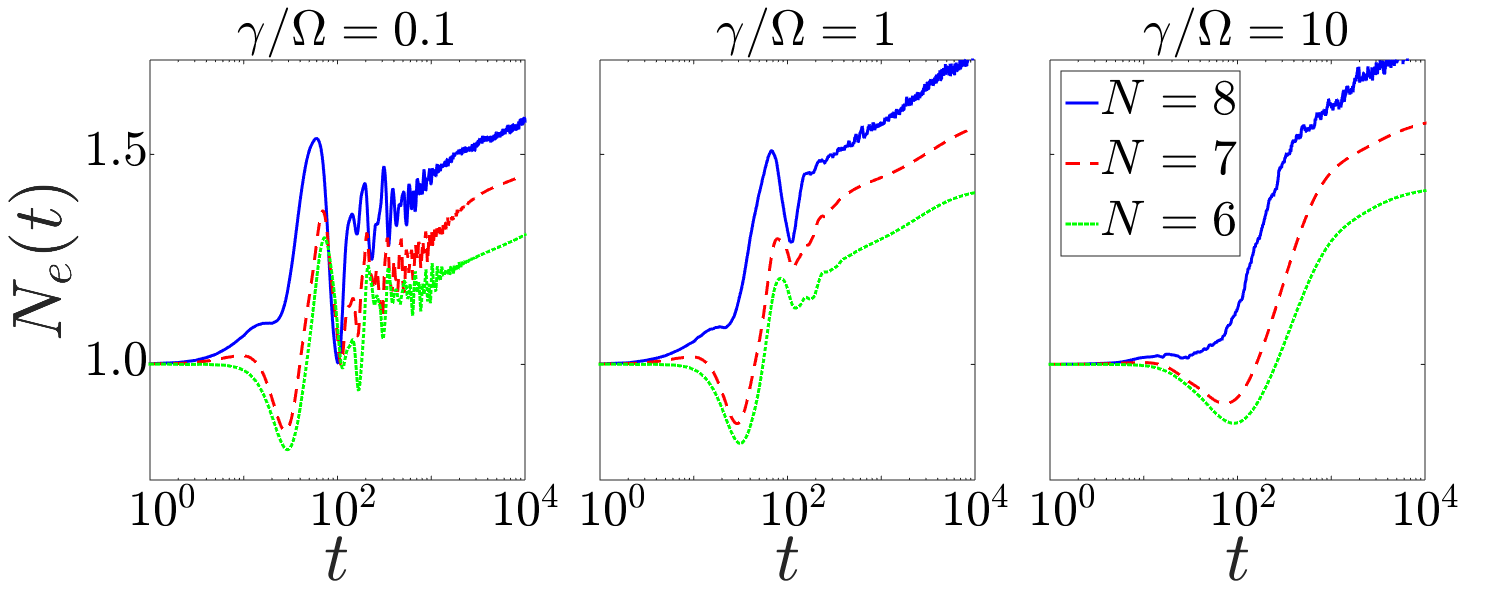}
\end{center}
\caption{ {\sf \bf Number of Rydberg excitations as a function of time for different system sizes.}
The number of excitations $N_e$ is displayed as a function of time for $\gamma/\Omega=0.1,1,10$ from left to right respectively. We considered the lengths $N=6,7,8$.}
\label{figApp}
\end{figure}
In this appendix we report on the qualitative algebraic increase in the number of Rydberg excitations described in Section 3 for different system sizes. 
Our results are displayed in Fig. \ref{figApp} for $\gamma/\Omega=0.1,1,10$.  
We considered a chain of length $N=6,7,8$, as larger sizes are numerically prohibitive. 
We find that this change in the system size does not alter the qualitative picture reported in the main text. Other ratios $\gamma/\Omega$ were considered leading to the same conclusions.

\end{document}